OPEN

# Entangling three qubits without ever touching

Pawel Blasiak[1,2]* & Marcin Markiewicz[3]

All identical particles are inherently correlated from the outset, regardless of how far apart their creation took place. In this paper, this fact is used for extraction of entanglement from independent particles unaffected by any interactions. Specifically, we are concerned with operational schemes for generation of all tripartite entangled states, essentially the GHZ state and the W state, which prevent the particles from touching one another over the entire evolution. The protocols discussed in the paper require only three particles in linear optical setups with equal efficiency for boson, fermion or anyon statistics. Within this framework indistinguishability of particles presents itself as a useful resource of entanglement accessible for practical applications.

> "*The whole is other than the sum of its parts.*"
> – *Aristotle (Metaphysics, Book 8)*

Indistinguishability brings a new quality into collective behaviour of quantum systems which manifests in novel correlation effects and subtle role of entanglement played in theoretical description. In the first quantisation formalism, the symmetrisation postulate requires that all identical particles are described by an entangled state from the very beginning. This in particular includes particles created independently in remote regions of the universe. Therefore, since *otherness* of entanglement manifests in non-local correlations[1], it should be in principle possible to observe these effects extracted directly from the symmetrized state of independent identical particles too[2–10]. In the second quantisation formalism, this entanglement is however masked by mode description which takes into account inability to directly address individual particles. Moreover, all experimental scenarios require some kind of interaction to unlock this potential, thereby blurring the origin of the observed phenomena (i.e., making unclear whether the correlation effects are due to the fundamental indistinguishability or just a consequence of interactions between the particles along the way). Is it thus possible to use this inherent form of entanglement in its pure form without calling for interaction to bring it into effect? In this paper we give a positive answer showing that entanglement due to indistinguishability is not just an artefact of the formalism but, on the contrary, can be turned into a useful resource accessible for practical purposes in protocols which circumvent the problem of particle interactions.

In order to be more precise we will distinguish between various kinds of interaction/correlation scenarios establishing entanglement between systems. A common intuition associates interaction event with a well defined space region in which particles or systems happen to be present at the same time. We shall call this requirement the *touching condition*. Then the interaction can have a typically *dynamical* character expressed by mixing terms in the Hamiltonian which couple respective modes of the system, e.g. like in description of fundamental interactions in particle physics or generation of entangled photons in spontaneous parametric down-conversion[11]. An alternative mechanism for correlating particles at the touching point is through the interference effects for identical particles, e.g. like the Pauli exclusion principle for fermions or bunching (anti-bunching) effect for bosons (fermions) impinging on a beam splitter[11,12]. This is a *kinematical* phenomenon which is determined by the commutation relations at the touching point with a different behaviour depending on the statistics of the particles involved.

In this paper, we go beyond the touching paradigm by considering situations in which particles *do not* meet at any point over the entire evolution, and yet correlation between particles is established. This can be realised in quantum optical frameworks by demanding spatial separation (or no-crossing) of the paths traversed by the particles for certain post-selected events. We shall call it the *interaction without touching* scenario. This striking idea was first proposed by B. Yurke and D. Stoler[13,14] who devised a scheme for generation of the Bell states and the GHZ state from two and three independent particles respectively. Conceptually, it plays with the naive intuition about what counts as an interaction process, since apparently spatial correlations can be established in a kinematical way without particles touching one another. All these scenarios rely entirely on particle *indistinguishability*, which thus presents itself as a useful resource of entanglement readily accessible for practical applications.

---

[1]Institute of Nuclear Physics Polish Academy of Sciences, PL-31342, Kraków, Poland. [2]City, University of London, London, EC1V OHB, UK. [3]Institute of Physics, Jagiellonian University, PL-30348, Kraków, Poland. *email: pawel. blasiak@ifj.edu.pl

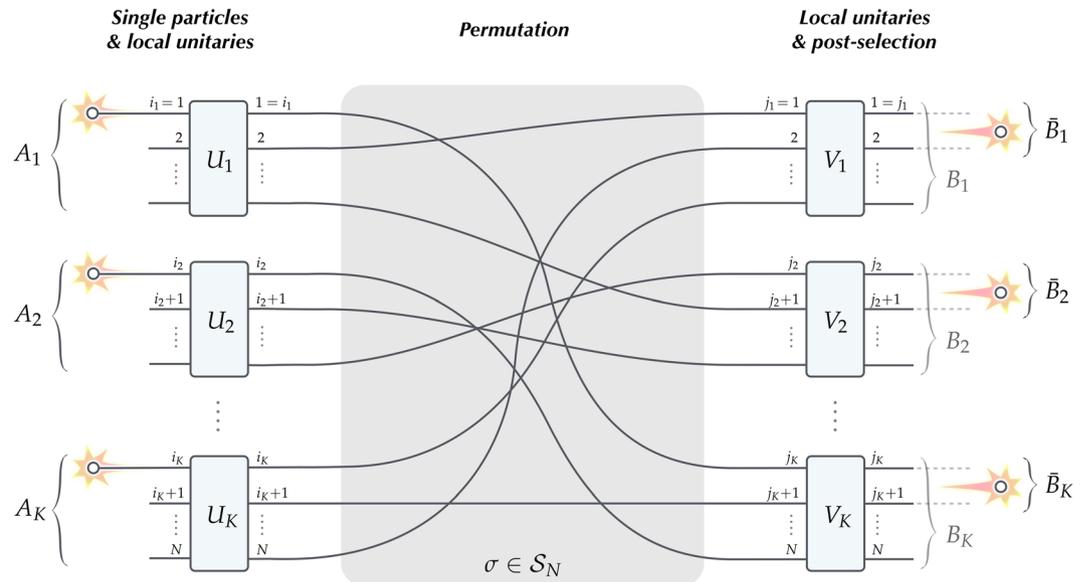

**Figure 1.** Interaction without touching scenario. Read from left to right, $K$ independent particles created in distant regions of space called subsystems $A_1, ..., A_K$ are transformed by local unitaries $U_k$ in the respective regions $A_k$ producing a product state of $K$ qudits (encoded as a superposition of paths grouped in $A_k$ for $k = 1, ..., K$)[15]. Then the paths are redirected to other space locations, following some permutation $\sigma \in S_N$, where the paths form $K$ groups of separated subsystems $B_1, ..., B_K$. The latter are further manipulated by local unitaries $V_k$ within the respective regions $B_k$. In the last step, the focus is on some chosen pairs of paths $\bar{B}_k \subset B_k$ picked out from each subsystem $B_k$ with the promise of post-selection which retains only the events with a single particle in each $\bar{B}_1, ..., \bar{B}_K$. This guarantees that such prepared (dual-rail) qubits $\bar{B}_k$ are well-defined. It is crucial to observe that post-selection and the specific geometry of the setups prevents the particles from touching one another over the entire evolution.

The problem posed in this paper is exploration of the extent to which extraction of entanglement is possible within the no-touching paradigm. More precisely, we consider the question of generation of all entangled states of a given (tripartite) type and the arrangements required for their construction. While for bipartite states the original proposal[13] provides the answer (two particles are enough), for tripartite entanglement the problem is already unsettled (only the construction of the GHZ state using three particles is known[14]). In this paper, we complete the task of generation of all tripartite entangled states by providing a simple scheme for generation of the W state using three independent particles in the no-touching scheme. The construction is preceded by a short discussion of a general interaction without touching scenario for dual-rail encoded qubits.

## Results

**Interaction without touching scenario.** Consider optical experiment with $N$ paths (or modes) which in the input and output are grouped into $K$ disjoint subsystems denoted by $A_1, ..., A_K$ and $B_1, ..., B_K$. Let the paths be labeled by consecutive integers and then we choose $A_k = \{i_k, i_k + 1, ...\}$ with $1 = i_1 < i_2 < ... < i_K < N$ and $B_k = \{j_k, j_k + 1, ...\}$ with $1 = j_1 < j_2 < ... < j_K < N$. Further, within each subsystem in the output we pick out pairs of paths $\bar{B}_k = \{j_k, j_k + 1\} \subset B_k$ which in the following will encode target qubits $\bar{B}_1, ..., \bar{B}_K$. In this paper we are interested in optical scenarios composed of five steps illustrated in Fig. 1:

(1) *Single particle* is injected into each group $A_1, ..., A_K$, say the first path $i_k$ in the respective subsystem $A_k$.
(2) *Local unitaries* are applied to each group $A_1, ..., A_K$ in the input, i.e. $U_k$ acts only on paths in subsystem $A_k$.
(3) *Permutation* of the paths in the circuit, i.e. rearranging paths according to some fixed permutation $\sigma \in S_N$.
(4) *Local unitaries* are applied to each group $B_1, ..., B_K$ in the output, i.e. $V_k$ acts only on paths in subsystem $B_k$.
(5) *Post-selection* on events with a single particle in each pair of paths (target qubits) $\bar{B}_1, ..., \bar{B}_K$ in the output.

In the optical context it is appropriate to think of the paths as spatially separated modes of the system with the particles described in the occupation number representation (i.e., the Fock space). Then the design of the circuit can be such that the subsystems $A_1, ..., A_K$ as well as $B_1, ..., B_K$ are located in separated space regions. In this sense the unitaries $U_1, ..., U_K$ and $V_1, ..., V_K$ are said to implement local transformations. The non-trivial part consists in rearranging the paths by permutation $\sigma \in S_N$ which is a physical transformation redirecting the paths to different locations (switching the fibres). We note that the latter introduces non-separability into the system, but crucially this can be implemented in a way that keeps all paths apart. (This point is important for the discussion focused on locality issues and the no-touching argument. For the pure task of entanglement generation in a system treated as a whole spatial separation of the modes is less important.)

In the following we adopt a naive point of view of particles traversing well-defined paths. This means that the particle may change its path only at the crossing points which correspond to the beam splitters implementing non-trivial unitaries on the respective modes. These are also the only points where the particles in different paths may touch. Therefore, after steps *(1)* and *(2)* we can be sure that in each group of paths $A_1, ..., A_K$ there is a single particle which has never touched the other ones. Clearly, permutation in step *(3)* does not introduce any crossing/touching too. Then, in step *(4)* we are back to the case of separate processing of a single particle within each subsystem $B_1, ..., B_K$. The latter is due to post-selection in *(5)* which retains only the events with a single particle in each $\tilde{B}_k \subset B_k$. This means that in the post-selected regime the no-touching condition is preserved throughout the whole procedure. (To put it differently, since the only crossing/touching points are associated either with the initial processing by local unitaries in $A_1, ..., A_K$ or local post-processing in $B_1, ..., B_K$, then the injection scheme prevents touching in the $A_k$'s and post-selection excludes touching in the $B_k$'s. Of course, processing in the $B_k$'s may involve crossing of the paths where two particles entering given $B_k$ can meet, but such events are rejected by the post-selection condition and hence irrelevant for the argument).

Note that treated collectively each group of paths $A_k$ with a single particle in it and local unitaries $U_k$ form a subsystem which encodes a qudit[15]. Therefore, by virtue of *(1)* and *(2)*, it is legitimate to think of the initial stage of this scenario as the representation of $K$ qudits $A_1, ..., A_K$. By the same token, post-selection *(4)* asserts presence of a single particle in each pair of paths $\tilde{B}_k$ and hence at the output subsystems $\tilde{B}_1, ..., \tilde{B}_K$ encode state of $K$ qubits (so called 'dual-rail' qubits). In a nutshell, the protocol prepares a state of $K$ dual-rail qubits represented in subsystems $\tilde{B}_1, ..., \tilde{B}_K$ from $K$ independent particles which have never touched along the way. It is interesting to ask whether this method is capable of generating all states in $\mathscr{H}^{\otimes K}$ where $\mathscr{H} = \mathbb{C}^2$ is a qubit.

**A note on notation.** Optical circuits analysed in this paper are constructed in such a way that at all times there is at most a single particle occupying each mode/path of the system. In the Fock space representation such states are combinations of the occupation number states with 0's and 1's, i.e.,

$$|. . . 10. . . 01. . . 10. . .\rangle = a_{l_1}^\dagger a_{l_2}^\dagger . . . a_{l_K}^\dagger |0\rangle, \tag{1}$$

with $l_1 < l_2 < ... < l_K$ specifying the occupied modes. In the following, we use the notation with creation operators which encodes particle statistics in the commutation relations. Since we are considering the no-touching scenarios which are arranged so that the particles are always in different modes/paths, then the only relevant commutator in the paper involves creation operators with different indices, i.e.,

$$a_k^\dagger a_l^\dagger = \pm\ a_l^\dagger a_k^\dagger \quad \text{for}\ k \neq l, \tag{2}$$

with the $\pm$ sign corresponding to the boson/fermion statistics respectively ('for anyons it gets replaced with the phase $e^{i\varphi}$'). This, in particular, means that bunching or anti-bunching effects play no role in these scenarios.

Note that the occupation number representation implicitly assumes *indistinguishability* of particles. This is crucial for the interference effects, which in the considered scenarios lead to entanglement in the output. For distinguishable particles these protocols give only classical correlations (see Methods section for discussion).

The dual-rail representation of a qubit consists in encoding of the Hilbert space $\mathscr{H} = \mathbb{C}^2$ as a superposition of single-particle states in a given pair of modes/paths. Accordingly, representation of a qubit in paths $\tilde{B}_k$ boils down to the following identification: $\alpha |\uparrow\rangle_k + \beta |\downarrow\rangle_k \equiv (\alpha\, a_{j_k}^\dagger + \beta\, a_{j_k+1}^\dagger) |0\rangle$ with unitaries implemented by local transformations (such as beam splitters and phase shifters) on the corresponding modes/paths $a_j^\dagger \to \sum_{l \in \tilde{B}_k} U_{lj}\, a_l^\dagger$ for $j \in \tilde{B}_k$ and measurement realised by particle detection[15]. This straightforwardly generalises to the dual-rail encoding of a general state of multiple qubits $\mathscr{H}^{\otimes K}$ in pairs of modes/paths, say $\tilde{B}_1, ..., \tilde{B}_K$, with the stipulation that in each pair only a single particle is present. In our scenarios this is guaranteed by the post-selection procedure which we denote by the arrow '$\leadsto$'. Note that the latter does not prevent from further processing of the system by local unitaries acting within each individual pair $\tilde{B}_1, ..., \tilde{B}_K$, since in that case post-selection can be deferred until all qubits will have been measured. In particular, correlation experiments with arbitrary measurements on qubits $\tilde{B}_1, ..., \tilde{B}_K$ can be performed in this way. This is consistent with the view that entanglement arises as an interference effect due to spatial overlap of the wave functions. Our setups have the advantage of separating this effect from the particle aspect of the description, i.e., in spite of the fact that the particles do not touch at any point, their wave functions eventually overlap in the detectors.

**Two notable examples.** Consider dual-rail encoding of two qubits $A_1 = \{1, 2\}$ and $A_2 = \{3, 4\}$ in distant space regions which transform via simple optical circuit into pair of qubits $\tilde{B}_1 = B_1 = \{1, 2\}$ and $\tilde{B}_2 = B_2 = \{3, 4\}$ in other spatially separated locations. See Fig. 2 (on the left) for explanation. Post-selection on the events with a single particle in each subsystem (target qubit) $\tilde{B}_1$ and $\tilde{B}_2$ prepares the following (unnormalised) state:

$$a_1^\dagger a_3^\dagger |0\rangle \xrightarrow{H,H} \frac{1}{2}\left(a_1^\dagger + a_2^\dagger\right)\left(a_3^\dagger + a_4^\dagger\right)|0\rangle$$
$$\xrightarrow{\sigma_{Bell}} \frac{1}{2}\left(a_1^\dagger + a_4^\dagger\right)\left(a_3^\dagger + a_2^\dagger\right)|0\rangle$$
$$\xrightarrow{post\text{-}select} \frac{1}{2}\left(a_1^\dagger a_3^\dagger + a_4^\dagger a_2^\dagger\right)|0\rangle, \tag{3}$$

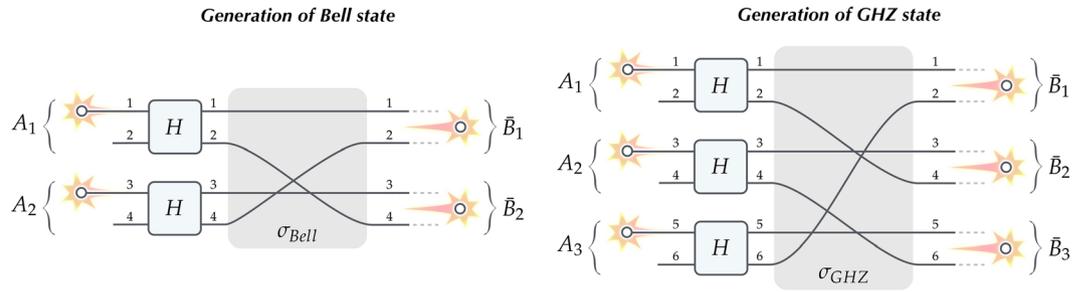

**Figure 2.** Two notable examples. Generation of the Bell state and the GHZ state in the no-touching scenario. Two (three) independent particles are processed without touching leading to the Bell (GHZ) state encoded in subsystems $B_1 = \tilde{B}_1$, $B_2 = \tilde{B}_2$ (and $B_3 = \tilde{B}_3$) in the output; see Eqs. (3) and (4). In both cases, local unitaries $H$ in the input are the Hadamard transforms and the permutations $\sigma_{Bell} \in S_4$ and $\sigma_{GHZ} \in S_6$ respectively $(1234) \mapsto (1432)$ for the Bell state and $(123456) \mapsto (143652)$ for the GHZ state.

which is the Bell state $\frac{1}{\sqrt{2}} \left( |\uparrow\uparrow\rangle \pm |\downarrow\downarrow\rangle \right)$ with the $\pm$ sign depending on the boson/fermion statistics of the particles (for anyons it gets replaced with the phase $e^{i\phi}$). The protocol can be generalised, by modifying beam splitters in the input and appending two beam splitters in the output, to generate *any* state of two qubits in $\mathcal{H} = \mathbb{C}^2 \otimes \mathbb{C}^2$. This is essentially a rewriting of the proposal by Yurke and Stoler[13] for entangling particles from two independent sources devised for testing Bell inequalities. In Methods section we explicitly show why the entire scheme fails to produce entangled states in the case of distinguishable particles.

Another example along the same lines is entangling three dual-rail qubits using three particles from independent sources in separated space regions $A_1 = \{1, 2\}$, $A_2 = \{3, 4\}$ and $A_3 = \{5, 6\}$; see Fig. 2 (on the right) for explanation. Upon post-selection on the events with a single particle in each subsystem (target qubit) $\tilde{B}_1 = B_1 = \{1, 2\}$, $\tilde{B}_2 = B_2 = \{3, 4\}$ and $\tilde{B}_3 = B_3 = \{5, 6\}$ the scheme leads to the following (unnormalised) state:

$$
\begin{aligned}
a_1^\dagger a_3^\dagger a_5^\dagger |0\rangle & \xrightarrow{H,H,H} & \frac{1}{2\sqrt{2}} \left( a_1^\dagger + a_2^\dagger \right) \left( a_3^\dagger + a_4^\dagger \right) \left( a_5^\dagger + a_6^\dagger \right) |0\rangle \\
& \xrightarrow{\sigma_{GHZ}} & \frac{1}{2\sqrt{2}} \left( a_1^\dagger + a_4^\dagger \right) \left( a_3^\dagger + a_6^\dagger \right) \left( a_5^\dagger + a_2^\dagger \right) |0\rangle \\
& \stackrel{post\text{-}select}{\rightsquigarrow} & \frac{1}{2\sqrt{2}} \left( a_1^\dagger a_3^\dagger a_5^\dagger + a_4^\dagger a_6^\dagger a_2^\dagger \right) |0\rangle,
\end{aligned}
\tag{4}
$$

which is the GHZ state $\frac{1}{\sqrt{2}} \left( |\uparrow\uparrow\uparrow\rangle \pm |\downarrow\downarrow\downarrow\rangle \right)$. In this case the result is insensitive to the boson/fermion statistics of the particles used in the protocol (for anyons due to two inversions the relative phase $e^{2i\phi}$ appears).

Both above examples demonstrate a practical realisation of the interaction without touching scenario with unitaries $H$ being the usual Hadamard transforms acting on the two paths entering the gate, i.e.,

$$
\begin{pmatrix} a_k^\dagger \\ a_l^\dagger \end{pmatrix} \xrightarrow{H} \frac{1}{\sqrt{2}} \begin{pmatrix} 1 & 1 \\ 1 & -1 \end{pmatrix} \begin{pmatrix} a_k^\dagger \\ a_l^\dagger \end{pmatrix}.
\tag{5}
$$

Let us note that post-selection in both protocols comes at a price of lowering the efficiency rate, i.e., generation of the Bell state in Eq. (3) succeeds with probability 1/2 and for the GHZ state in Eq. (4) the probability is 1/4.

**Generation of the W state without touching.** There are essentially only two interesting entangled states of three qubits, i.e., the GHZ state and the W state. Any other tripartite entangled state in $\mathcal{H} = \mathbb{C}^2 \otimes \mathbb{C}^2 \otimes \mathbb{C}^2$ can be produced from one of these two states by stochastic local operations and classical communication (SLOCC) splitting tripartite entanglement into two disjoint classes[16]. We have seen above that the GHZ state can be obtained from three independent particles in the no-touching scenario. In the following, we complete the task of generation of tripartite entanglement by constructing the W state from three particles without touching.

Consider the following 7-path scenario, illustrated in Fig. 3, in which the input and output paths are grouped into three spatially separated subsystems: $A_1 = B_1 = \{1, 2\}$, $A_2 = B_2 = \{3, 4, 5\}$ and $A_3 = B_3 = \{6, 7\}$. For the target dual-rail qubits we choose: $\tilde{B}_1 = \{1, 2\}$, $\tilde{B}_2 = \{3, 4\}$ and $\tilde{B}_3 = \{6, 7\}$. Let the initial state of the system consist of three particles described as $a_1^\dagger a_3^\dagger a_6^\dagger |0\rangle$. In the first step, local unitaries $H$ are the Hadamard transforms on subsystems $A_1, A_3$ and the unitary $U$ on subsystem $A_2$ is chosen in such a way to produce the following superposition:

$$
a_3^\dagger \xrightarrow{U} \frac{1}{\sqrt{5}} \left( \sqrt{2}\, a_3^\dagger + a_4^\dagger + \sqrt{2}\, a_5^\dagger \right).
\tag{6}
$$

Further, we switch the paths according to the permutation $\sigma_W \in S_7$ given by $(1234567) \mapsto (1324765)$. In the final step, only paths 3 and 5 in subsystem $B_2$ are unitarily processed by the Hadamard transform $H$ and path 4 is left unaffected. As per usual, the protocol assumes post-selection which consists in retaining only those events when a single particle is present in each target qubit $\tilde{B}_1, \tilde{B}_2$ and $\tilde{B}_3$ in the output (which does not preclude further processing within each $\tilde{B}_1, \tilde{B}_2, \tilde{B}_3$).

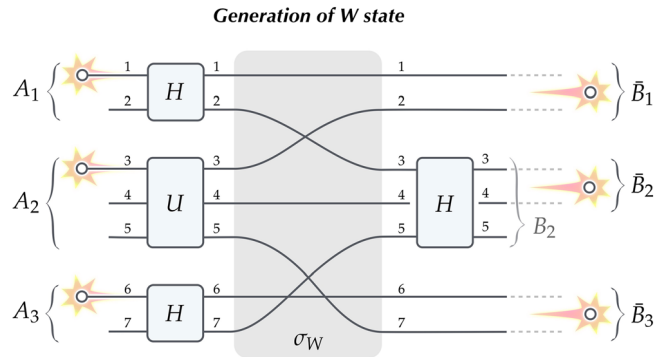

**Figure 3.** Generation of the W state without touching. Three independent particles injected into separated subsystems $A_1$, $A_2$ and $A_3$ produce the W state encoded in subsystems $\tilde{B}_1$, $\tilde{B}_2$ and $\tilde{B}_3$ (dual-rail qubits). The protocol follows the interaction without touching scenario with post-selection in subsystems $\tilde{B}_1$, $\tilde{B}_2$ and $\tilde{B}_3$, where $H$ are the usual Hadamard transforms Eq. (5) and $U$ is the unitary specified in Eq. (6). See details in the text.

Taking all steps together evolution of the input state goes as follows:

$$
\begin{aligned}
a_1^\dagger a_3^\dagger a_6^\dagger \,|0\rangle \;&\xrightarrow{H,U,H}\; \frac{1}{2\sqrt{5}}\left(a_1^\dagger + a_2^\dagger\right)\left(\sqrt{2}\,a_3^\dagger + a_4^\dagger + \sqrt{2}\,a_5^\dagger\right)\left(a_6^\dagger + a_7^\dagger\right)|0\rangle \\
&\xrightarrow{\sigma_W}\; \frac{1}{2\sqrt{5}}\left(a_1^\dagger + a_3^\dagger\right)\left(\sqrt{2}\,a_2^\dagger + a_4^\dagger + \sqrt{2}\,a_7^\dagger\right)\left(a_6^\dagger + a_5^\dagger\right)|0\rangle \\
&\xrightarrow{H}\; \frac{1}{2\sqrt{5}}\left(a_1^\dagger + \frac{a_3^\dagger + a_5^\dagger}{\sqrt{2}}\right)\left(\sqrt{2}\,a_2^\dagger + a_4^\dagger + \sqrt{2}\,a_7^\dagger\right)\left(a_6^\dagger + \frac{a_3^\dagger - a_5^\dagger}{\sqrt{2}}\right)|0\rangle \\
&\xrightarrow{post\text{-}select}\; \frac{1}{2\sqrt{5}}\left(a_1^\dagger a_4^\dagger a_6^\dagger + a_1^\dagger a_7^\dagger a_3^\dagger + a_3^\dagger a_2^\dagger a_6^\dagger\right)|0\rangle,
\end{aligned}
\tag{7}
$$

which is the W state $\frac{1}{\sqrt{3}}\left(|\!\uparrow\downarrow\uparrow\rangle \pm |\!\uparrow\uparrow\downarrow\rangle \pm |\!\downarrow\uparrow\uparrow\rangle\right)$ with the $\pm$ sign depending on the boson/fermion statistics of the particles used in the protocol (for anyons it should be replaced with the phase $e^{i\phi}$). Due to post-selection the procedure succeeds with probability $\left(\sqrt{3}\,/\,2\sqrt{5}\right)^2 = 15\%$. The efficiency rate and simplicity of the setup should be compared with generation of the W state with touching for polarisation-encoded qubits, e.g., in Refs. [5,17–20].

## Discussion

Generation of entanglement, besides importance for practical implementations of quantum processing tasks, touches upon the fundamental question of the origins of non-locality and quantum correlations. Interestingly, entanglement can be established in scenarios which do not require prior interaction between the particles, understood as an event happening at some well-defined touching point (i.e., in situations when the specific geometry of the setups prevents the particles from touching one another over the entire evolution). In this paper, we have discussed interaction without touching scenarios for dual-rail qubits and described explicit protocols for generation of arbitrary tripartite entanglement (essentially the GHZ and the W state) from *indistinguishable particles* within the no-touching paradigm. Notably, the construction of the W state is to our knowledge the only example in the literature not requiring auxiliary particles, cf. Refs. [5,17–20].

A similar idea of processing without touching is present in the so called entanglement swapping protocol [21]. Note, however, that the latter requires explicitly entangled particles to start off, and hence it should be seen as a powerful state manipulation rather than state generation technique. Another relevant example is the ingenious proposal by L. Hardy for testing local realism in the overlapping interferometers [22]. In that scheme the entangled state is designed through post-selection due to the interaction event in the touching point (electron-positron annihilation [22] or photon bunching effect [23]). In this respect, scenarios considered in this paper are conceptually different, since entanglement is obtained *'almost for nothing'*. More specifically, (*i*) no prior entanglement is assumed except for that granted by the fundamental principle of particle indistinguishability, and (*ii*) no conditioning on any kind of prior interaction/touching along the way is required which is guaranteed by the non-crossing geometry of the setups.

Interest in entanglement for systems of identical particles comes from potential applications in quantum information processing. A notable effort in clarifying the role of indistinguishability as a resource has been recently made in Refs. [2–10]. In particular, in a series of works Refs. [4,5,8] it was shown that bringing identical particles into the same spatial location may serve as an entangling operation when local counting (post-selection) of particles is made. In this paper, the crucial difference lays in the use of dual-rail encoding of qubits which allows for design of circuits with the no-touching property guaranteed by post-selection. This eliminates all sorts of effects like dynamical interaction between the particles or kinematical bunching (anti-bunching) phenomena. Formally, in the second quantisation formalism it boils down to the irrelevance of the creation-annihilation commutator for the description of such designed protocols. It shows that entanglement can be generated without bringing any of the particles into the same spatial location, thus rendering schemes without touching different from those mentioned above. On top of the conceptual differences there is also a practical advantage: (*i*) no auxiliary particles are needed (just two particles for bipartite entanglement and three particles for tripartite entanglement) and (*ii*) efficiency does not depend on the statistics of the particles used in the protocols (creation-annihilation commutator is irrelevant in the post-selected regime).

Apart from the pure task of entanglement generation, our work sheds a new light on some fundamental issues concerning quantum optical interferometry. Firstly, it surpasses the very reductionist approach to quantum interferometry, which describes all phenomena in terms of mode excitations and detector clicks. Instead we consider a more realistic approach with the explicit concept of a particle as a single entity which can be localized in a path of the interferometer. The crucial aspect of this viewpoint is that this sort of weak realism does not lead to any contradiction, as opposed to the Bell-type and contextual scenarios. On the contrary, one can say in old terms that the effect of no-touching is an *element of reality*, since conditioning on the outcomes of the experiment (post-selection) we can say with certainty that the particles have never met during the experiment. The second foundational issue concerns the problem of locality of operations in interferometric setups. The no-touching scheme illustrates a new kind of duality present in these frameworks. Namely, operations perfectly global from the mode perspective (permutation of modes) turn out to be perfectly local in the weakly-realistic particle picture, which is reflected by the fact that the post-selected particles never localize in the same point of space. This property is absent in optical schemes using bunching (anti-bunching), mode mixing or any sort of referential systems.

To conclude, the protocols discussed in the paper treat particle *indistinguishability* as a genuine resource of entanglement which can be successfully unlocked into a useful form that can be directly observed in the correlation experiments; see Refs. [13,14] for the seminal observation, Refs. [24,25] for some recent realisations with electrons and Refs. [26,27] for photonic implementations. We leave open an interesting question whether these scenarios can be generalised to produce all multipartite entangled states for more than three qubits.

## Methods

### Failure of the schemes for distinguishable particles.

In order to better appreciate particle indistinguishability for generation of entanglement in the considered scenarios, we will explicitly examine the simplest case of the Bell state in Eq. (3), Fig. 2 on the left, and see how it performs when the particles are distinguishable. Let us distinguish the particles by using different letters $a^\dagger$ and $b^\dagger$ for the creation operators describing particles entering subsystem $A_1$ and $A_2$ respectively. Then, evolution of the system leads to the following state:

$$
\begin{aligned}
a_1^\dagger b_3^\dagger \,|0\rangle \quad &\xrightarrow{H,H} \quad \tfrac{1}{2}\big(a_1^\dagger + a_2^\dagger\big)\big(b_3^\dagger + b_4^\dagger\big)|0\rangle \\
&\xrightarrow{\sigma_{Bell}} \quad \tfrac{1}{2}\big(a_1^\dagger + a_4^\dagger\big)\big(b_3^\dagger + b_2^\dagger\big)|0\rangle \\
&\xrightsquigarrow{post\text{-}select} \quad \tfrac{1}{2}\big(a_1^\dagger b_3^\dagger + a_4^\dagger b_2^\dagger\big)|0\rangle.
\end{aligned}
\tag{8}
$$

Despite a similar form the resulting state should not be confused with the state in Eq. (3). In order to see the difference it is however not enough to make a measurement in the computational basis (i.e., placing detectors in each path in the output), since in such an experiment the statistics will be the same as for the state in Eq. (3). One needs to make a general correlation experiment by performing unitaries $V_1$ and $V_2$ on subsystems $B_1$ and $B_2$ respectively (implemented by phase shifters and beam splitters) and then detect particles in each path (these two steps correspond to local change of measurement basis). For simplicity we take real unitaries in the form:

$$
V_1 = \begin{pmatrix} p & q \\ q & -p \end{pmatrix} \quad \text{and} \quad V_2 = \begin{pmatrix} s & t \\ t & -s \end{pmatrix},
\tag{9}
$$

with $p^2 + q^2 = 1$, $s^2 + t^2 = 1$, which yield the following evolution in the case of indistinguishable and distinguishable particles respectively:

$$
\begin{aligned}
\text{Eq. (3)} \quad \xrightarrow{V_1,V_2} \quad &\tfrac{1}{2}\big((p\,a_1^\dagger + q\,a_2^\dagger)(s\,a_3^\dagger + t\,a_4^\dagger) + (t\,a_3^\dagger - s\,a_4^\dagger)(q\,a_1^\dagger - p\,a_2^\dagger)\big)|0\rangle \\
= \quad &\tfrac{1}{2}\big((ps \pm tq)\,a_1^\dagger a_3^\dagger + (pt \mp sq)\,a_1^\dagger a_4^\dagger \\
&+ (qs \mp tp)\,a_2^\dagger a_3^\dagger + (qt \pm sp)\,a_2^\dagger a_4^\dagger\big)|0\rangle,
\end{aligned}
\tag{10}
$$

$$
\begin{aligned}
\text{Eq. (8)} \quad \xrightarrow{V_1,V_2} \quad &\tfrac{1}{2}\big((p\,a_1^\dagger + q\,a_2^\dagger)(s\,b_3^\dagger + t\,b_4^\dagger) + (t\,a_3^\dagger - s\,a_4^\dagger)(q\,b_1^\dagger - p\,b_2^\dagger)\big)|0\rangle \\
= \quad &\tfrac{1}{2}\big(ps\,a_1^\dagger b_3^\dagger + pt\,a_1^\dagger b_4^\dagger + qs\,a_2^\dagger b_3^\dagger + qt\,a_2^\dagger b_4^\dagger + tq\,a_3^\dagger b_1^\dagger \\
&- tp\,a_3^\dagger b_2^\dagger - sq\,a_4^\dagger b_1^\dagger + sp\,a_4^\dagger b_2^\dagger\big)|0\rangle,
\end{aligned}
\tag{11}
$$

with the $\pm$ signs in Eq. (10) depending on the boson or fermion statistics of the particles used in the protocol (for anyons it should be replaced with the phase $e^{i\phi}$). Note that due to indistinguishability some interference in Eq. (10) occurs, while in Eq. (11) all the terms remain distinct. Now, we detect particles in each path and perform the usual Bell correlation test between subsystems $B_1$ and $B_2$, i.e., calculate the correlation function $\langle B_1 B_2 \rangle_{V_1 V_2}$ where $B_1 = \pm 1$ corresponds to the particle being detected respectively in path 1 or 2 of subsystem $B_1$ and similarly $B_2 = \pm 1$ corresponds to the particle being detected respectively in path 3 or 4 of subsystem $B_2$. From Eqs. (10) and (11) upon normalisation, we get:

 6

$$\text{Indistinguishable:} \quad \langle B_1 B_2 \rangle_{V_1 V_2} = \frac{1}{2}\left((ps \pm tq)^2 + (qt \pm sp)^2 - (pt \mp sq)^2 - (qs \mp tp)^2\right)$$
$$= \cos(\theta_1 \mp \theta_2), \tag{12}$$

$$\text{Distinguishable:} \quad \langle B_1 B_2 \rangle_{V_1 V_2} = (ps)^2 + (tq)^2 - (pt)^2 - (qs)^2$$
$$= \cos(\theta_1) \cdot \cos(\theta_2), \tag{13}$$

where in the last equality the following parametrisation is used: $p = \cos\frac{\theta_1}{2}, q = \sin\frac{\theta_1}{2}$ and $s = \cos\frac{\theta_2}{2}, t = \sin\frac{\theta_2}{2}$. Clearly, for distinguishable particles the correlation function factorizes and hence Bell inequalities are obeyed (i.e., there is no entanglement in that case), while for indistinguishable particles the correlation function takes the familiar non-factorizable form expected from the Bell state $\frac{1}{\sqrt{2}}(|\uparrow\uparrow\rangle \pm |\downarrow\downarrow\rangle)$ (i.e., the state is entangled).

From the above discussion it is interesting to observe that in these scenarios for indistinguishable particles non-local correlations arise due to the interference of quantum amplitudes in the measuring devices (understood as particle detection preceded with a unitary transformation). This is consistent with a heuristic principle saying that quantum interference occurs only when paths traversed by the particles are indistinguishable. In other words, if one can (even in principle) distinguish the path each particle has taken, then only classical correlations are obtained. From diagrams in Figs. 1, 2 and 3 it is easy to see that detection of the particle after the unitary transformation merging two paths entering subsystem $B_k$ does not tell anything about which path the particle came from only if the particles are indistinguishable. Only in this case interference effects lead to correlations beyond the Bell bound. For distinguishable particles quantum interference does not occur which obstructs generation of entanglement in the considered scenarios.

## References

1. Bell, J. S. *Speakable and unspeakable in quantum mechanics*. (Cambridge University Press, 1987).
2. Wiseman, H. M. & Vaccaro, J. A. Entanglement of Indistinguishable Particles Shared between Two Parties. *Phys. Rev. Lett.* **91**, 097902 (2003).
3. Killoran, N., Cramer, M. & Plenio, M. B. Extracting Entanglement from Identical Particles. *Phys. Rev. Lett.* **112**, 150501 (2014).
4. Lo Franco, R. & Compagno, G. Quantum entanglement of identical particles by standard information-theoretic notions. *Sci. Rep.* **6**, 20603 (2016).
5. Bellomo, B., Lo Franco, R. & Compagno, G. N identical particles and one particle to entangle them all. *Phys. Rev. A* **96**, 022319 (2017).
6. Compagno, G., Castellini, A. & Lo Franco, R. Dealing with indistinguishable particles and their entanglement. *Philos. Trans. Roy. Soc. A* **376**, 20170317 (2018).
7. Lo Franco, R. & Compagno, G. Indistinguishability of Elementary Systems as a Resource for Quantum Information Processing. *Phys. Rev. Lett.* **120**, 240403 (2018).
8. Castellini, A., Bellomo, B., Compagno, G. & Lo Franco, R. Activating remote entanglement in a quantum network by local counting of identical particles. *Phys. Rev. A* **99**, 062322 (2019).
9. Bouvrie, P. A., Valdés-Hernández, A., Majtey, A. P., Zander, C. & Plastino, A. R. Entanglement generation through particle detection in systems of identical fermions. *Ann. Phys.* **383**, 401 (2017).
10. Bose, S. & Home, D. Duality in Entanglement Enabling a Test of Quantum Indistinguishability Unaffected by Interactions. *Phys. Rev. Lett.* **110**, 140404 (2013).
11. Pan, J.-W. *et al.* Multiphoton entanglement and interferometry. *Rev. Mod. Phys.* **84**, 777 (2012).
12. Tichy, M. C. Interference of identical particles from entanglement to boson-sampling. *J. Phys. B: At. Mol. Opt. Phys.* **47**, 103001 (2014).
13. Yurke, B. & Stoler, D. Bell's-inequality experiments using independent-particle sources. *Phys. Rev. A* **46**, 2229 (1992).
14. Yurke, B. & Stoler, D. Einstein-Podolsky-Rosen Effects from Independent Particle Sources. *Phys. Rev. Lett.* **68**, 1251 (1992).
15. Reck, M., Zeilinger, A., Bernstein, H. J. & Bertani, P. Experimental Realization of Any Discrete Unitary Operator. *Phys. Rev. Lett.* **73**, 58 (1994).
16. Dür, W., Vidal, G. & Cirac, J. I. Three qubits can be entangled in two inequivalent ways. *Phys. Rev. A* **62**, 062314 (2000).
17. Eibl, M., Kiesel, N., Bourennane, M., Kurtsiefer, C. & Weinfurter, H. Experimental Realization of a Three-Qubit Entangled W State. *Phys. Rev. Lett.* **92**, 077901 (2004).
18. Walther, P., Resch, K. J. & Zeilinger, A. Local Conversion of Greenberger-Horne-Zeilinger States to Approximate W States. *Phys. Rev. Lett.* **94**, 240501 (2005).
19. Tashima, T. *et al.* Local Transformation of Two Einstein-Podolsky-Rosen Photon Pairs into a Three-Photon W State. *Phys. Rev. Lett.* **102**, 130502 (2009).
20. Krenn, M., Hochrainer, A., Lahiri, M. & Zeilinger, A. Entanglement by Path Identity. *Phys. Rev. Lett.* **118**, 080401 (2017).
21. Żukowski, M., Zeilinger, A., Horne, M. A. & Ekert, A. K. "Event-Ready-Detectors" Bell Experiment via Entanglement Swapping. *Phys. Rev. Lett.* **71**, 4287 (1993).
22. Hardy, L. Quantum mechanics, local realistic theories, and Lorentz-invariant realistic theories. *Phys. Rev. Lett.* **68**, 2981 (1992).
23. Irvine, W. T. M., Hodelin, J. F., Simon, C. & Bouwmeester, D. Realization of Hardy's Thought Experiment with Photons. *Phys. Rev. Lett.* **95**, 030401 (2005).
24. Dasenbrook, D. *et al.* Single-electron entanglement and nonlocality. *New J. Phys.* **18**, 043036 (2016).
25. Neder, I. *et al.* Interference between two indistinguishable electrons from independent sources. *Nature* **448**, 333 (2007).
26. Li, X.-M., Yang, M., Paunković, N., Li, D.-C. & Cao, Z.-L. Entanglement swapping via three-step quantum walk-like protocol. *Phys. Lett. A* **381**, 3875 (2017).
27. Kim, Y.-S. *et al.* Informationally symmetrical Bell state preparation and measurement. *Opt. Express* **26**, 29539 (2018).

## Acknowledgements

We thank M. Karczewski, Y.-S. Kim, M. Krenn, P. Kurzyński, M. Kuś, R. Lo Franco, M. Yang, J. Vaccaro, H. Wiseman and M. Żukowski for helpful comments. PB acknowledges support from the Polish National Agency for Academic Exchange in the Bekker Scholarship Programme. MM acknowledges the National Science Centre (Poland), through Grant No. 2015/16/S/ST2/00447, within the FUGA 4 project for postdoctoral training.

## Author contributions

Both authors researched and wrote the paper.

## Competing interests

The authors declare no competing interests.

## Additional information

**Correspondence** and requests for materials should be addressed to P.B.

**Reprints and permissions information** is available at www.nature.com/reprints.

**Publisher's note** Springer Nature remains neutral with regard to jurisdictional claims in published maps and institutional affiliations.